\documentclass{rspublic}
\usepackage[pdftex,final]{graphicx}
\usepackage{url}
\usepackage[round,colon,authoryear]{natbib}
\usepackage{multirow}

\usepackage[pdftex]{color}

\providecommand{\pdiff}[2]{\frac{\partial #1}{\partial #2}}

\providecommand{\abs}[1]{\left\lvert#1\right\rvert}
\providecommand{\transpose}[1]{\ensuremath{#1^\mathrm{T}}}

\providecommand{\figref}[1]{Figure~\ref{#1}}
\providecommand{\secref}[1]{Section~\ref{#1}}

\renewcommand{\eqref}[1]{[\textbf{\ref{#1}}]}
\providecommand{\eqrefs}[1]{[\textbf{\ref{#1}}]}

\providecommand{\secrange}[2]{Sections~\ref{#1}--\ref{#2}}

\begin{document}

\title{Optimal strategies for throwing accurately}

\author{Madhusudhan Venkadesan$^\dagger$ and L.\ Mahadevan$^\ddag$}
\affiliation{School of Engineering \& Applied Sciences,
Harvard University, Cambridge, MA 02138, USA}
\footnotetext[1]{mvenkadesan@gmail.com}
\footnotetext[2]{lm@seas.harvard.edu}
\maketitle

\begin{abstract}{throwing, noise propagation,  projectile motion}
Accuracy of throwing in games and sports is governed by how errors at projectile release are propagated by flight dynamics. To address the question of what governs the choice of throwing strategy, we use a simple model of throwing with an arm modelled as a hinged bar of fixed length that can release a projectile at any angle  and angular velocity. We show that the amplification of deviations in launch parameters from a one parameter family of solution curves is quantified by the largest singular value of an appropriate Jacobian. This allows us to predict a preferred throwing style in terms of this singular value, which itself depends on target location and the target shape. Our analysis also allows us to characterize the trade-off between speed and accuracy despite not including any effects of signal-dependent noise. Using nonlinear calculations for propagating finite input-noise, we find that an underarm throw to a target leads to an undershoot, but an overarm throw does not. Finally, we consider the limit of the arm-length vanishing, i.e.\ shooting a projectile, and find that the most accurate shooting angle bifurcates as the ratio of the relative noisiness of the initial conditions crosses a threshold.
\end{abstract}

\section{Introduction}

What governs the selection of a style when throwing a ball into a bin, say overarm versus underarm? At first the question seems ill-posed
since of the arbitrarily large number of possible choices and variables: people prefer overarm over underarm or vice versa depending on
age, gender, culture, training etc. However, the question becomes better posed if we consider just the physics of  projectile release with a
prescribed velocity and launch angle to reach the target. This is because there is a  one parameter family of solutions; for almost any given launch 
angle, we can prescribe a launch velocity. How then might we choose from this family of trajectories ? Since any errors introduced in the 
initial conditions are not uniformly amplified by all trajectories,  a natural possibility is that we ought to choose that trajectory that least 
amplifies any initial errors. This leads us naturally to the subject of error propagation and amplification in dynamical systems with
antecedents that go back a century or more \citep{Poincare1912,Hopf1934} but continue to have implications for prediction and 
decision making.   Here, we consider the simple task of throwing for two reasons.  First, the task decouples the internal (neural) decision 
making from the physics of projectiles, i.e.\ the task decouples planning from feedback control. In the absence of wind, there is no noise 
during the projectile flight, only amplification of initial noise. Second, this sets the stage for understanding strategy and learning using a 
trial-to-trial iterative process of error estimation and correction.

Not surprisingly, throwing accuracy has been studied by many others in various contexts such as clinical applications 
\citep{Timmann1999,Powls1995}, human evolution \citep{Wood2007,Watson2001,Westergaard2000,Chowdhary1999}, sports 
\citep{Gablonsky2005,Smeets2002,Dupuy2000,Brancazio1981,Tan1981,Okubo2006}, and human motor control 
\citep{Hore1996,Kudo2000,Cohen2009}. However, none of them address the choice of throwing style, or present an analysis of error 
propagation through projectile dynamics in the presence of an arm, which we consider here.

In \secref{sec:FactorsThatAffectThrowingStyle} we outline the various possible physical and biological influences on throwing strategy. 
\secref{sec:ModelOfTheThrowingTask} develops our model of throwing along with the solution for a perfect strike, while in
\secref{sec:QuantificationOfErrorAmplification} we use a linearization of the mapping between release parameters 
and projectile landing location to quantify error amplification. In \secref{sec:velocityAccuracyTradeOff} 
we calculate the dependence of error amplification on throwing speed.
\secref{sec:PropagatingDistributionsWithNonInfinitesimalVariance} presents the a fully nonlinear calculations for 
propagating noise in release parameters with assumed distributions, and compares the linear and 
nonlinear predictions using a specific numerical example. \secref{sec:ShootingZeroArmLength} analyzes the limit of vanishing arm length, i.e.\ the 
problem of shooting a projectile and calculatesthe optimal shooting angle as a function of the relative noise level in shooting angle 
versus speed. We conclude in \secref{sec:Discussion} with a comparison of our findings with previous results, and some implications of our results.
  
\section{Factors that affect throwing style}
\label{sec:FactorsThatAffectThrowingStyle}
The choice of strategy for any motor task is influenced by a number of physical and biological factors. For throwing, the physics of 
projectile flight is affected by the projectile's shape, air drag, spin, wind, and so on, which in turn affects error propagation by flight 
dynamics. The decision of how to release a projectile is determined by a number of biological factors including muscle strength, skeletal 
joint limits, sensory feedback, memory, and learning. 

Here, we only study the case of windless, drag-free point-like projectile with no spin. The target can itself have many variations such as its 
shape, orientation, size, possibility of rebounds, a backboard (like in basketball), and so on. We consider two target geometries: (i) an 
upward facing, horizontal line-like target of prescribed small extent, and (ii) a circular target of prescribed small extent.

The complex articulation of the human arm allows it to control four independent launch parameters in the plane of projectile flight, for 
example, the location and velocity of the projectile. Here we model the arm much more simply as a hinged rigid bar connecting the 
shoulder to the hand, so that it can only control two independent launch parameters. Although the neural controller can control any two 
independent launch parameters, say the speed and angle of release, or the speed and timing of release of the projectile, and so on, we 
use the angle the arm makes with the horizontal ($\phi$), and the angular velocity ($\Omega$). We also disregard any strength 
differences between throwing styles. We note that prescribing an alternate set of variables is not completely equivalent to our choice, 
since the Jacobian of the transformation will necessarily play an important role in determining how errors are amplified. This implies
an important role for the choice of representation of the control variables for the motor task, although here we will limit ourselves to just
a single convenient one given above.

The variability and correlations in the release parameters ($\phi$, $\Omega$) depend on the detailed properties of the neuromuscular 
system. For simplicity, here we use uncorrelated noise in $\phi$ and $\Omega$. For the linear analysis we assume infinitesimally small 
noise in $\phi$ and $\Omega$ and therefore need no further assumptions about the specific distribution. 

\begin{figure}
	\centering
		\includegraphics{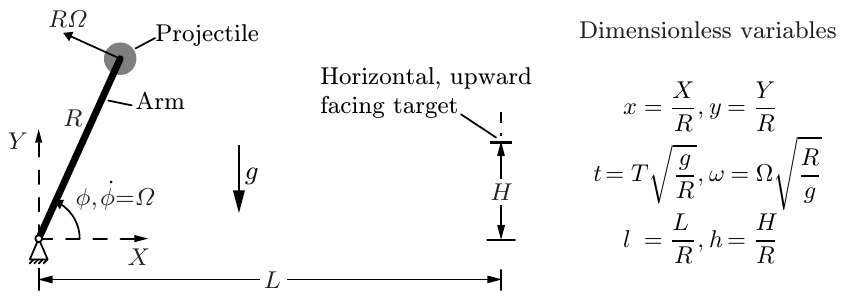}
	\caption{Model of the arm used to investigate how errors in throwing are propagated by projectile mechanics. The arm is modelled 
	as a hinged rod of fixed length whose angle and angular velocity can be fully controlled. Using this `super-human' arm model that 
	has no limits on strength or coordination ability allows us to focus only on the projectile mechanics without consideration of ability to 
	control the arm. $R$ is the arm's length, $\phi$ and $\Omega$ are the angle measured from the $+X$ axis and angular velocity of 
	the arm, respectively. $L$ and $H$ specify the target coordinates, and $g$ is gravitational acceleration.}
	\label{fig:fig2-armmodel}
\end{figure}

\section{Model of throwing}
\label{sec:ModelOfTheThrowingTask}
With the simplifications outlined above, we model the arm as a hinged bar of fixed length with independent control over the arm's release 
angle ($\phi$) and angular velocity ($\Omega$) as shown in \figref{fig:fig2-armmodel}. For a drag-free point-like projectile moving in a 
uniform gravitational field, its trajectory $(X(T), Y(T))$ as a function of time $T$ measured following release is obtained by solving the 
initial value problem,
\begin{subequations}\label{eqn:flightequations}
\begin{align}
	&\ddot{X}(T) = 0,\ \dot{X}(0) = -R\Omega\sin\phi,\ X(0) = R\cos\phi \label{eqn:Xeqns}\\
	&\ddot{Y}(T) = -g,\ \dot{Y}(0) = R\Omega\cos\phi,\ Y(0) = R\sin\phi \label{eqn:Yeqns}
\end{align}
\end{subequations}
where, $X$ and $Y$ are measured relative to the arm's pivot, $R$ is the arm length, and $g$ is the acceleration due to gravity. We use 
the arm's length $R$ and its natural frequency $\sqrt{g/R}$ to define the following dimensionless variables,
\begin{equation} \label{eqn:nonD}
	x = \frac{X}{R},\quad y = \frac{Y}{R},\quad t = T\sqrt{\frac{g}{R}},\quad \omega = \Omega\sqrt{\frac{R}{g}}
\end{equation}
Then the dimensionless equations for the projectile to land on a horizontally oriented, planar target (like a bin) located at a scaled height 
$h = H/R$, and distance $l=L/R$ are,
\begin{subequations}\label{eqn:xyt}
\begin{align}
	x(t,\phi,\omega) &= \cos\phi - t\omega\sin\phi = l \label{eqn:x(t,phi,omega)}\\
	y(t,\phi,\omega) &= \sin\phi + t\omega\cos\phi - \frac{1}{2}t^2 = h \label{eqn:y(t,phi,omega)}
\end{align}
\end{subequations}

By solving equation~\eqref{eqn:y(t,phi,omega)} for $t$, one equation with three unknowns, and using the condition $\dot{y}(t) = \omega
\cos\phi - t \le 0$ for the projectile to strike the upper face of the target, we obtain a solution surface $t_h(\phi,\omega)$. By substituting 
$t=t_h(\phi,\omega)$ in equation~\eqref{eqn:x(t,phi,omega)} we also obtain an equation for the horizontal landing location of the projectile 
when it strikes the plane of the target.
\begin{subequations}\label{eqn:thxh}
\begin{align}
	t_h(\phi,\omega) &= \omega\cos\phi + \sqrt{\omega^2\cos^2\phi - 2(h-\sin\phi)}\label{eqn:th}\\
	x_h(\phi,\omega) &= \cos\phi - \omega\sin\phi\left(\omega\cos\phi + \sqrt{\omega^2\cos^2\phi-2(h-\sin\phi)}\right)\label{eqn:xh}
\end{align}
\end{subequations}
Note the necessary condition that $\omega^2 \ge 2(h-\sin\phi)/\cos^2\phi$, i.e.\ a minimum launch velocity is needed to reach the target's 
plane when $h\ge\sin\phi$.

\begin{figure}
	\centering
		\includegraphics{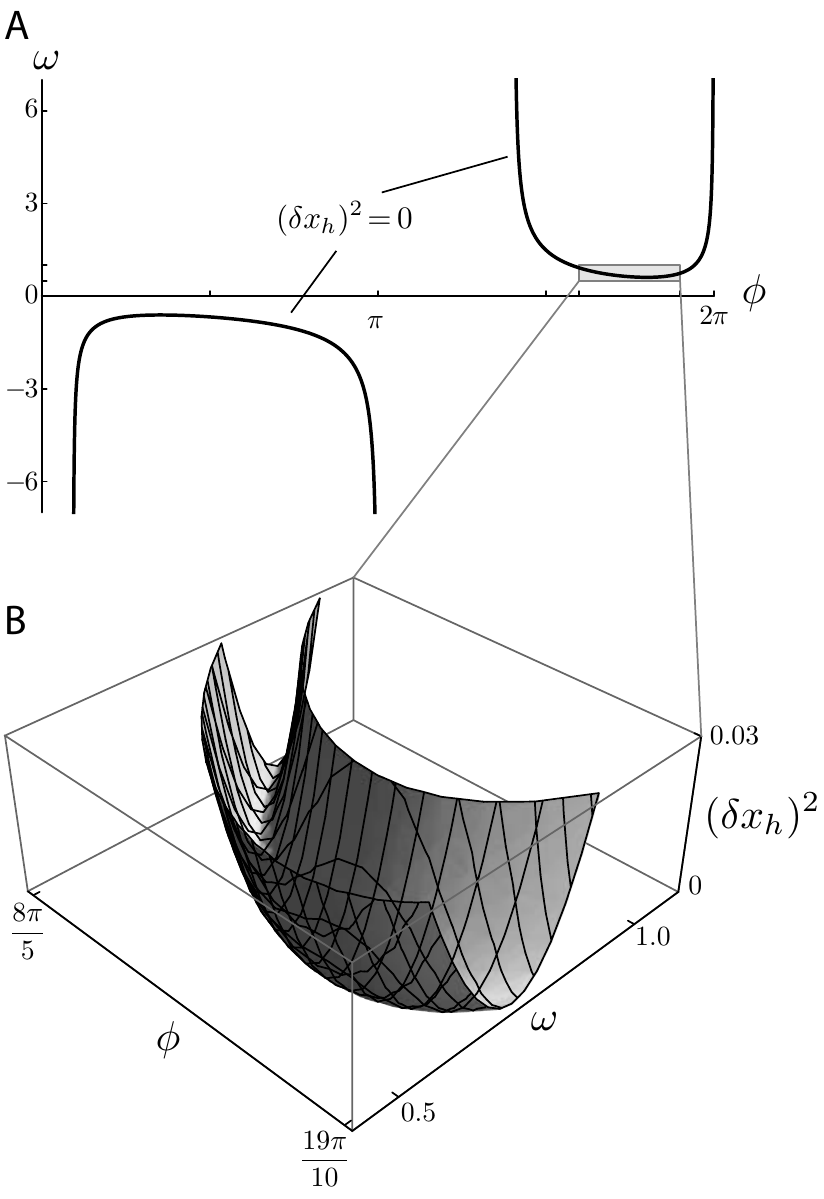}
	\caption{One-dimensional curve of parameters for an accurate strike, and error amplification. \textbf{a.}\ For a given target location 
	($l=1.5, h=-1.5$) there is a curve of parameters $(\phi,\omega_0(\phi))$ for exactly striking the target. The portion with positive $
	\omega$ is the underarm throw according to our sign conventions and $\omega$ is negative for the overarm throw. There are some 
	angles of release ($\phi$) for which is it impossible to strike the target as seen by the non-overlapping region between the underarm 
	and overarm throws. \textbf{b.}\ Deviations away from the curve leads to an error $\delta x_h$, and we plot the squared error. Local 
	error amplification near $(\phi,\omega_0(\phi))$ is quantified by the maximal curvature that varies with $\phi$.}
	\label{fig:planar_VelAng-QuadPit_raster}
\end{figure}

Then we find the conditions for exactly striking the target by solving equations~\eqref{eqn:xyt} simultaneously, or by setting $x_h(\phi,
\omega)=l$ in equation~\eqref{eqn:xh} to obtain a one parameter family of solutions  
\begin{align}
	\omega_0(\phi) &= \frac{\cos\phi-l}{\sin\phi\sqrt{\frac{2}{\sin\phi}\left(1-l\cos\phi-h\sin\phi\right)}}\label{eqn:omega0}
\end{align}
 as shown, for example, in \figref{fig:planar_VelAng-QuadPit_raster}a.

\section{Quantification of error amplification}
\label{sec:QuantificationOfErrorAmplification}
When the release parameters $(\phi,\omega)$ deviate from the one-dimensional solution curve ($\phi,\omega_0(\phi)$), the projectile 
misses the target  leading to an error $\delta x_h = x_h(\phi,\omega) - l$. To quantify 
the amplification of small launch errors in the throw, we linearize $x_h(\phi,\omega)$ in the neighbourhood of the curve ($\phi,
\omega_0(\phi)$) to obtain the amplification of `input errors' $\transpose{(\delta\phi\ \delta\omega)}$ to the `output/target error' $\delta 
x_h$.
\begin{subequations}\label{eqn:Jerrderivation}
\begin{align}
&\delta x_h(\phi,\omega) \approx J_{\text{err}}(\phi)
				\begin{pmatrix}
						\delta\phi\\
						\delta\omega
				\end{pmatrix} \label{eqn:dxlin}\\
&\text{where}\ J_{\text{err}}(\phi) =	\begin{pmatrix}
								\pdiff{x_h}{\phi}& \pdiff{x_h}{\omega}
																\end{pmatrix}\bigg\vert_{\omega = \omega_0(\phi)} 
																\label{eqn:Jerr}
\end{align}
\end{subequations}

The amplification of errors as a function of $\phi$ is quantified by the largest singular value ($\lambda(\phi)$) of $J_{\text{err}}(\phi)$
(The explicit expressions for  $J_\text{err}(\phi)$ and $\lambda(\phi)$ are given in \ref{sec:LinearizationForThrowingWithAnArm}.).
Because there is a one-dimensional curve ($\phi,\omega_0(\phi)$) where $\delta x_h = 0$, $J_{\text{err}}$ will be rank deficient, i.e.\ it 
has a zero singular value and an associated non-trivial null-space, namely, the tangent to the curve ($\phi,\omega_0(\phi)$). 

By comparing the lowest order Taylor series expansion of $(\delta x_h)^2$ given by
\begin{equation}\label{eqn:dxnormexpression}
	(\delta x_h)^2 = 
						\transpose{\begin{pmatrix}\delta\phi\\ \delta\omega \end{pmatrix}}
							J_\text{err}^\text{T} J_\text{err}
							\begin{pmatrix}\delta\phi\\ \delta\omega \end{pmatrix}
								 = \frac{1}{2}
										\transpose{\begin{pmatrix}\delta\phi\\ \delta\omega \end{pmatrix}}
										H_\text{err}
										\begin{pmatrix}\delta\phi\\ \delta\omega \end{pmatrix}
\end{equation}
with equation~\eqref{eqn:dxlin}, we see that the quantity $2 \lambda^2$ is equal to the largest eigenvalue of the Hessian ($H_\text{err}$) 
of the output variance ($\delta x_h^2$) and thus the maximal principal curvature of the surface $\delta x_h^2(\phi, \omega_0(\phi)) = 0$ 
shown in \figref{fig:planar_VelAng-QuadPit_raster}b, leading to a geometric interpretation of error amplification.

\begin{figure}
	\centering
		\includegraphics{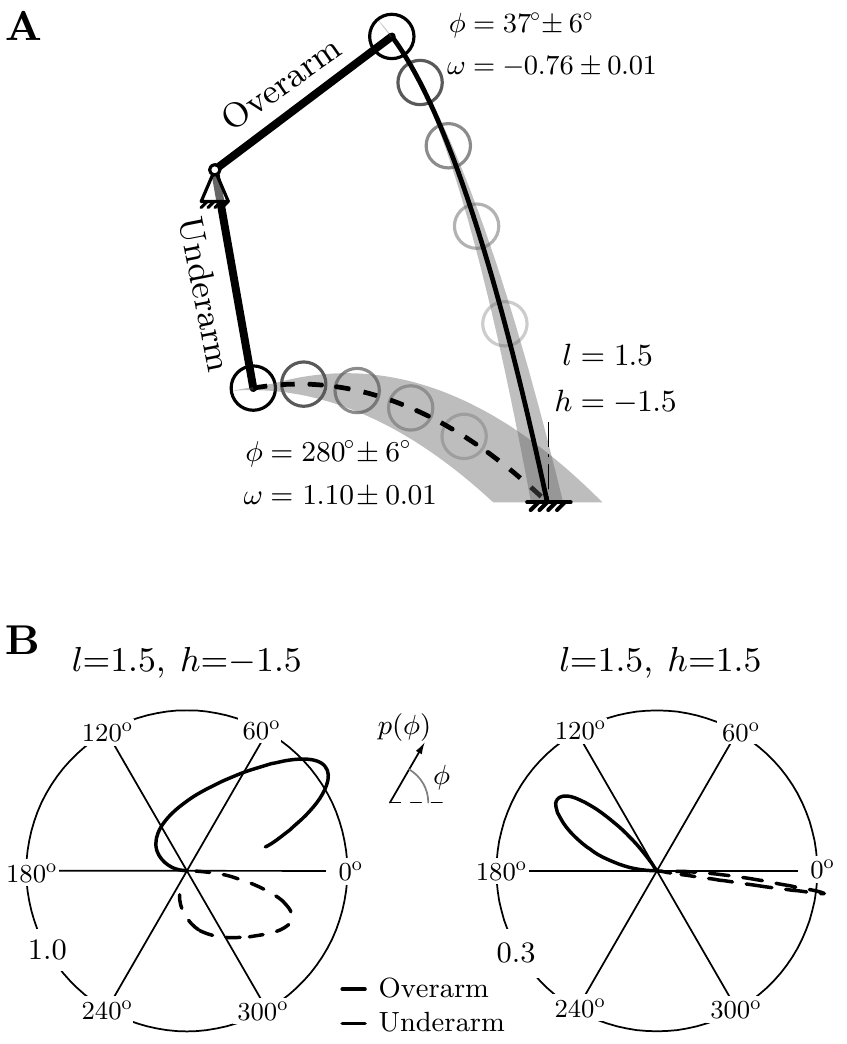}
	\caption{ Errors in releasing the projectile are amplified differently for different throwing strategies, depending on the target location and planning errors. \textbf{a.} This figure shows a comparison between a specific overarm (solid) and underarm (dashed) throw by the simple hinged arm. 
	Throws where $\omega_0(\phi) \ge 0$ are called `underarm' and those with $\omega_0(\phi) < 0$ are called `overarm'. For the 
	same initial errors in angle $\delta\phi$ and angular velocity $\delta\omega$ of the arm at release, the output error $\delta x_h$ for 
	the overarm strategy is lesser than the underarm strategy. We quantify this error amplification using the 
	non-zero singular value $\lambda(\phi)$ of the map $J_\text{err}:(\delta\phi,\delta\omega)^\text{T}\mapsto\delta x_h$. \textbf{b.} 
	Polar plots of accuracy $p(\phi) = 1/\lambda(\phi)$ as a function of arm angle at release $\phi$ for two different targets. If the choice of release angle is close to optimal, the underarm throw is clearly superior for the 
	target above the pivot, and the overarm strategy is better for the target below the pivot. But, in right polar plot, a small (but non-infinitesimal) deviation from the 
	optimum will cause a drastic deterioration in accuracy for the underarm throw, although it is superior in terms of its maximum.}
	\label{fig:numerical_example_polar_plots}
\end{figure}

To quantify the relative grading of strategies, we start with a simple numerical example shown in
\figref{fig:numerical_example_polar_plots}a which compares two throwing strategies---overarm (red) and underarm 
(green). Equal errors in release parameters amplify differently for different strategies, because $\lambda$ for the overarm throw is smaller 
than for the underarm throw for that target location. The envelope of trajectories, and hence landing locations of the projectile, was found 
by integrating forward a large number (100) trajectories starting from initial conditions in a box $(\phi\pm\delta\phi,\omega\pm\delta
\omega)$. The nominal values for $\phi$ and $\omega$ for the two throwing styles were chosen as the `optimal' strategy in the sense we 
define below.

Since the amplification of input error is directly proportional to $\lambda$, a natural metric to grade strategies for a given target ($l,h$) is 
the reciprocal of $\lambda$, which we will henceforth refer to as `throw accuracy', or simply `accuracy',
\begin{equation}\label{eqn:probofthrow}
	p(\phi) = \frac{1}{\lambda(\phi)},\quad \phi\in\{\phi':\delta x_h(\phi',\omega_0(\phi')) = 0\}
\end{equation}
In \figref{fig:numerical_example_polar_plots}b we show a polar plot with polar-angle $\phi$ and polar-radius $p(\phi)$, for two different 
target locations---one below and another above the arm's pivot. Recall that $\omega_0(\phi) \ge 0$ for underarm and $\omega_0(\phi) < 
0$ for overarm. For infinitesimal errors in release parameters, the best underarm throw is clearly superior for the target above the pivot, 
and the overarm strategy is better for the target below the pivot. If one were not a perfect planner, i.e.\ one does not choose the throwing 
angle $\phi$ that corresponds to the maximum, the answer is no longer as clear. In \figref{fig:numerical_example_polar_plots}b for 
example, the accuracy of the underarm throw in the right polar plot has a very narrow peak, i.e.\ small deviations in planning the angle of 
release from the optimum cause a drastic deterioration in accuracy. Therefore, although the underarm throw is superior for small planning 
errors, it is not robust to large planning errors. By accounting for both the peak and width of $p(\phi)$ we predict the fraction of overarm throws for a given target, and compared it against data in a separate publication \citep{Venkadesana}. 

\begin{figure}
	\centering
		\includegraphics{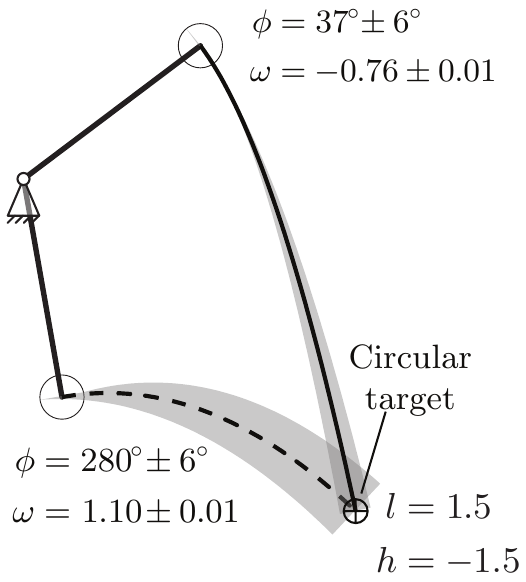}
	\caption{Errors for a circular target depend on the angle of approach of the projectile as seen in the numerical example shown above. This is because, error is defined as the distance of nearest approach.}
	\label{fig:circular_results_raster}
\end{figure}

All of the above calculations are for a flat, horizontal target, like a bin. However, if the target had a different shape, then target error will no longer be the horizontal deviation $\delta x_{h}$. For example, consider a ball-like circular target, where the target error will be the distance of nearest approach (\figref{fig:circular_results_raster}). Unlike a flat target (\figref{fig:numerical_example_polar_plots}A), the locus of nearest approach points for a circular target depends on the initial conditions (\figref{fig:circular_results_raster}). These differences in error propagation naturally bear out as sensitivity of the optimal throwing strategy to the shape of the target \citep{Venkadesana}. 

\section{Speed vs. accuracy}
\label{sec:velocityAccuracyTradeOff}
\begin{figure}
	\centering
		\includegraphics{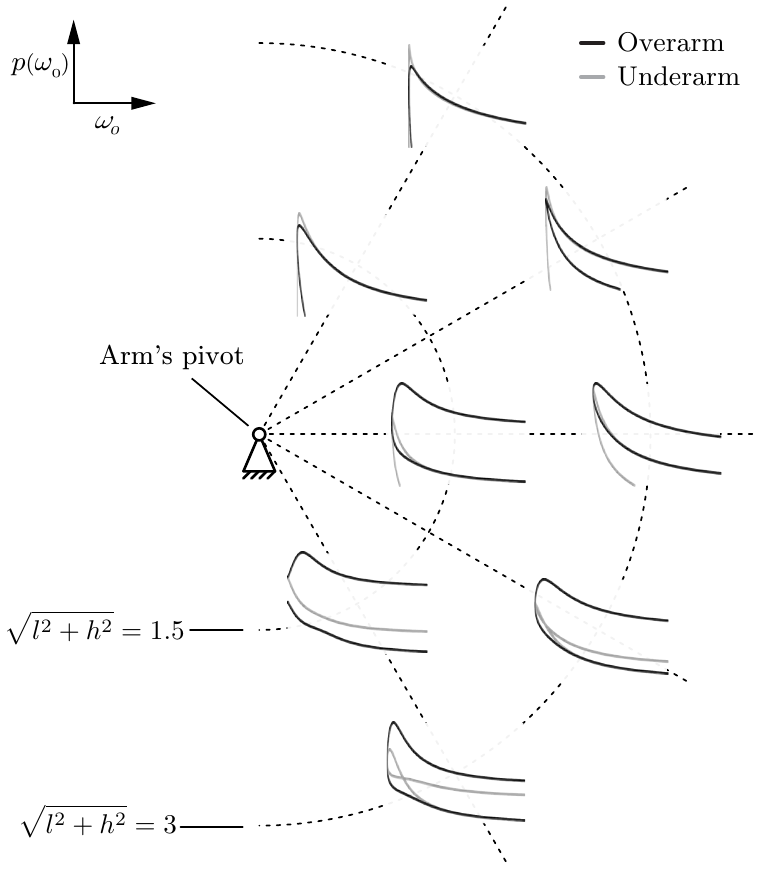}
	\caption{Speed-accuracy trade-off for various locations of the target. Each sub-plot shows $p(\omega_0)$ as a function of $
	\omega_0$ for a different target location. One could hit a given target with many possible velocities by changing the angle of release. 
	However, each launch velocity (and hence, each launch angle) has an associated error amplification quantified by $p = 1/\lambda$. 
	As seen in this figure, slower throws are typically more accurate than faster ones according to our model, much like the classic speed-accuracy trade-off that is seen in human motor behaviour. It is noteworthy that for some target 
	locations, our model predicts that the most accurate throw is at speeds slight higher than $\omega_\text{min}$ the smallest possible 
	speed to reach the target. This feature is more pronounced for nearer targets that are below the arm's pivot. Note that for $
	\omega_0 > \omega_\text{min}$, there are 4 different ways to hit the target, a shallow or a high throw and an underarm or overarm 
	throw. This figure shows $p(\omega_0)$ for each of these 4 throwing strategies by restricting our attention to $\omega_0 \in 
	[\omega_\text{min},3.54]$. For a human arm of 1 m length, $\omega=3.54$ corresponds to a launch speed of $\approx 11.1 
	\text{m/s}\ \approx 40 \text{km/h}$. We also find that when higher speed throws are necessary, a shallow overarm throw is more 
	accurate than any other strategy.}
	\label{fig:fig4-velocityaccuracyfig}
\end{figure}

We have so far presented accuracy $p$ as a function of release angle $\phi$ in the neighbourhood of the curve of accurate strike solutions $\omega_{0}(\phi)$. In order to understand if there is a relation between speed and accuracy, we re-parameterize $p$ as a function of $\omega$  by inverting the function $\omega_{0}(\phi)$ as shown in \figref{fig:fig4-velocityaccuracyfig} 
\footnote{We note that the function $\omega=\omega_{0}(\phi)$ is not bijective (\figref{fig:planar_VelAng-QuadPit_raster}a), and generally, the pre-image of every $\omega$ consists of four distinct values of $\phi$ corresponding to a high/shallow, overarm/underarm throw (four curves at each location in \figref{fig:fig4-velocityaccuracyfig}).}.  We see that slower throws are typically more accurate than faster ones in our model---the classic trade-off between speed and accuracy that is observed in human motor behaviour  \citep{Fitts1954,Meyer1990,Kerr1977,Harris1998,Harris2006,Todorov2004,Etnyre1998}. Such speed-accuracy trade-offs in biological systems has so far been explained as the result of signal-dependent and structured covariance of input noise at the level of muscles \citep{Harris1998,Harris2006,Todorov2004}. Our model however, has no strength limitations, and the input errors are additive, uncorrelated and independent of posture/velocity. Yet, we see the emergence of a speed-accuracy trade-off dictated only by projectile dynamics, and is thus a surprise.

For every target location, there exists a value $\omega_\text{min}(l,h)$, which is the smallest velocity needed to reach the target. For 
faster throwing velocities $\omega > \omega_\text{min}$, there are four distinct throws that hit the target --- overarm/underarm, shallow/
high throw. We find that the shallow overarm throw is most accurate for higher speeds, particularly for targets at or below the arm's pivot 
(equivalent to the shoulder). However, for slower speeds, the underarm throw is marginally more accurate than overarm for targets above 
the arm's pivot.

\section{Propagating distributions with non-infinitesimal variance}
\label{sec:PropagatingDistributionsWithNonInfinitesimalVariance}
We next relax the assumption of infinitesimally small input errors that was needed for the linearized analyses presented thus far. We introduce finite noise at the input side using a joint probability density function (PDF) $f_{[\phi,\omega]}$ 
associated with the noise in $\phi$ and $\omega$. The horizontal distance $x_{h}$ at which the projectile lands is given by the non-invertible, surjective function $x_h(\phi,\omega)$, which transforms $f_{[\phi,\omega]}$ into the PDF $f_{x_h}$  associated with $x_h$. For a fixed $\phi$, $f_{[\phi,\omega]}
(\phi,\omega_0(\phi,l))$ is transformed into $f_{x_h}(l)$ by the Jacobian of $\omega_0$ with respect to $l$. However, for a given target at 
distance $l$ and height $h$, there is a curve of release parameters $(\phi,\omega_0(\phi,l))$ for accurately striking the target. Therefore, 
integrating over this solution curve and using \eqref{eqn:omega0} for $\omega_0$, we find $f_{x_h}$ as,
\begin{subequations}
\begin{align}
	f_{x_h}(l) &= \frac{\int\limits_{\Phi}{f_{[\phi,\omega]}(\phi',\omega_0(\phi',l)) \abs{\pdiff{\omega_0(\phi',l)}{l}}\, d\phi'}}{\int\limits_{-
	\infty}^{\infty}{\int\limits_{\Phi}{f_{[\phi,\omega]}(\phi',\omega_0(\phi',l)) \abs{\pdiff{\omega_0(\phi',l)}{l}}\, d\phi'\, d l}}}\label{eqn:fxh}\\
	\pdiff{\omega_0(\phi,l)}{l} &= \frac{\csc ^2\phi (4 h\sin\phi + 2 l\cos\phi + \cos 2\phi - 3)}{4\sqrt{2} (\cot\phi+\csc\phi-h-l)^{3/2}}
	\label{eqn:Domega0Dl}
\end{align}
\end{subequations}
where $\Phi =\{\phi:x_h(\phi,\omega_0(\phi,l)) = l\}$ is typically disjoint and depends on the location of the target as enumerated in 
\ref{sec:FeasibleArmAngles}. To ensure that $f_{x_h}$ is a probability density with total area equal to one, we normalize \eqref{eqn:fxh} 
\footnote{There is `leakage' because $x_h(\phi,\omega)$ is not a real number for some combinations $(\phi,\omega)$ 
when the projectile fails to reach the plane of the target, and thus normalization becomes necessary.}.

To follow the implications of the fully nonlinear calculation for error amplification, we choose the simple example of throwing a projectile into a bin. For this example, we assume uncorrelated noise in $\phi$ and $\omega$, $f_{[\phi,\omega]} = f_\phi f_\omega$, where $f_\phi$ 
is a von Mises distribution. Because $\phi$ is a periodic variable, we use a von Mises distribution \citep{vonMises1918,Mardia2000}, which is the circular analogue of a Gaussian distribution. The PDF of a von Mises distribution with mean $\mu$ and concentration $\kappa$ ($1/\kappa$ is analogous to the variance of a normal distribution) is given by,
\begin{equation}\label{eqn:vonMisesPDF}
	v(x,\mu,\kappa) = \frac{e^{\kappa \cos(x-\mu)}}{2\pi I_{0}(\kappa)}
\end{equation}
where $I_{0}(\kappa)$ is the modified Bessel function of order 0. For $f_{\omega}$, we use a one-sided truncated Gaussian in 
order to restrict $\omega$ to a single throwing style at a time, i.e.\ $\omega\ge0$ or $\omega<0$ for underarm and overarm, respectively. The PDF of a truncated Gaussian variable $x$ with mean $\mu$ and standard deviation $\sigma$ that is truncated at $a$ and $b$ is given by,
\begin{equation}\label{eqn:TruncatedGaussianX}
g(x,\mu,\sigma,a,b) = \begin{cases}
 \sqrt{\frac{2}{\pi}}\frac{\frac{1}{\sigma}e^{-\frac{(x-\mu )^2}{2 \sigma ^2}}}{ 
   \left(\text{erf}\left(\frac{b-\mu }{\sigma\sqrt{2}}\right)
   -\text{erf}\left(\frac{a-\mu }{\sigma\sqrt{2}}\right)\right)} & \text{if}\ a\leq
   x\leq b \\
 0 & \text{Otherwise}
\end{cases}
\end{equation}					
where erf is the Gauss error function. Therefore, the distributions we use for numerical examples are,
\begin{subequations}\label{eqn:f_phi-omega}
\begin{align}
	f_{[\phi,\omega]}(\phi',\omega') &= f_\phi(\phi')\, f_\omega(\omega') \label{eqn:f_phi,omega}\\
	f_{\phi}(\phi') &= v(\phi',\phi^\text{optim}_{(\centerdot)},1/\sigma_\phi^{2}),\ (\centerdot)= \text{over or under} \label{f_phi}\\
	f_\omega(\omega') &= \begin{cases}
									g(\omega',\omega_0(\phi^\text{optim}_\text{over}),\sigma_\omega,-\infty,0) & \text{for 
									overarm}\\
									g(\omega',\omega_0(\phi^\text{optim}_\text{under}),\sigma_\omega,0,\infty) & \text{for 
									underarm}
								\end{cases} \label{eqn:f_omega}
\end{align}
\end{subequations}
where $\sigma$ is the standard deviation, and $\phi^\text{optim}$ denotes the optimal throwing angle such that the linear amplification $
\lambda(\phi^\text{optim})$ is the minimum for the chosen throwing style (\eqref{eqn:singular value formula} and Sections~
\ref{sec:QuantificationOfErrorAmplification}).

We now perform linear and nonlinear analyses for a specific numerical example, a bin that is 2 arm lengths away, and 1 arm length below the shoulder, i.e.\ $l=2$, $h=-1$. For this bin, we use the linear error amplification $\lambda$ to find the angle $\phi$ and speed $\omega$ for the optimal overarm and underarm throws. The input distributions for the nonlinear calculation are centred at the optimal release parameters found from this linear analysis. The variance of the distributions are arbitrarily chosen, but are equal for both throwing styles. Table~\ref{tab:paramvalues} lists the numerical values used in this example. Dimensional values are shown for a one meter long arm.

\begin{table}
\caption{Parameters values used in nonlinear calculations for bin at $l=2$, $h=-1$.}
\label{tab:paramvalues}
\centering
\begin{tabular}{|c|c|c|c|c|c|c|c|}
\hline 
\multicolumn{4}{|c|}{Overarm} & \multicolumn{4}{c|}{Underarm}\\
\hline
\multicolumn{2}{|c|}{$f_{\phi}$} & \multicolumn{2}{c|}{$f_{\omega}$} & \multicolumn{2}{c|}{$f_{\phi}$} & \multicolumn{2}{c|}{$f_{\omega}$}\\
\hline 
$\phi^\text{optim}_\text{over} $ & $\sigma_\phi$ & $\omega_0(\phi^\text{optim}_\text{over})$ & $\sigma_\omega$ & $\phi^\text{optim}_\text{under} $ & $\sigma_\phi$ & $\omega_0(\phi^\text{optim}_\text{under})$ & $\sigma_\omega$\\
\hline 
1.08 & 0.09 & -1.19 & 0.18 & 5.72 & 0.09 & 1.01 & 0.18\\
\hline
\multirow{2}{*}{$62^{\circ}$} & \multirow{2}{*}{$5^{\circ}$} & 3.72 m/s & 0.56 m/s & \multirow{2}{*}{$-33^{\circ}$} & \multirow{2}{*}{$5^{\circ}$} & 3.17 m/s & 0.56 m/s\\
 & & 13.4 km/hr & 2 km/hr & & & 11.4 km/hr & 2 km/hr\\
\hline
\end{tabular}
\end{table}

\begin{figure}
	\centering
		\includegraphics[width=1.0\textwidth]{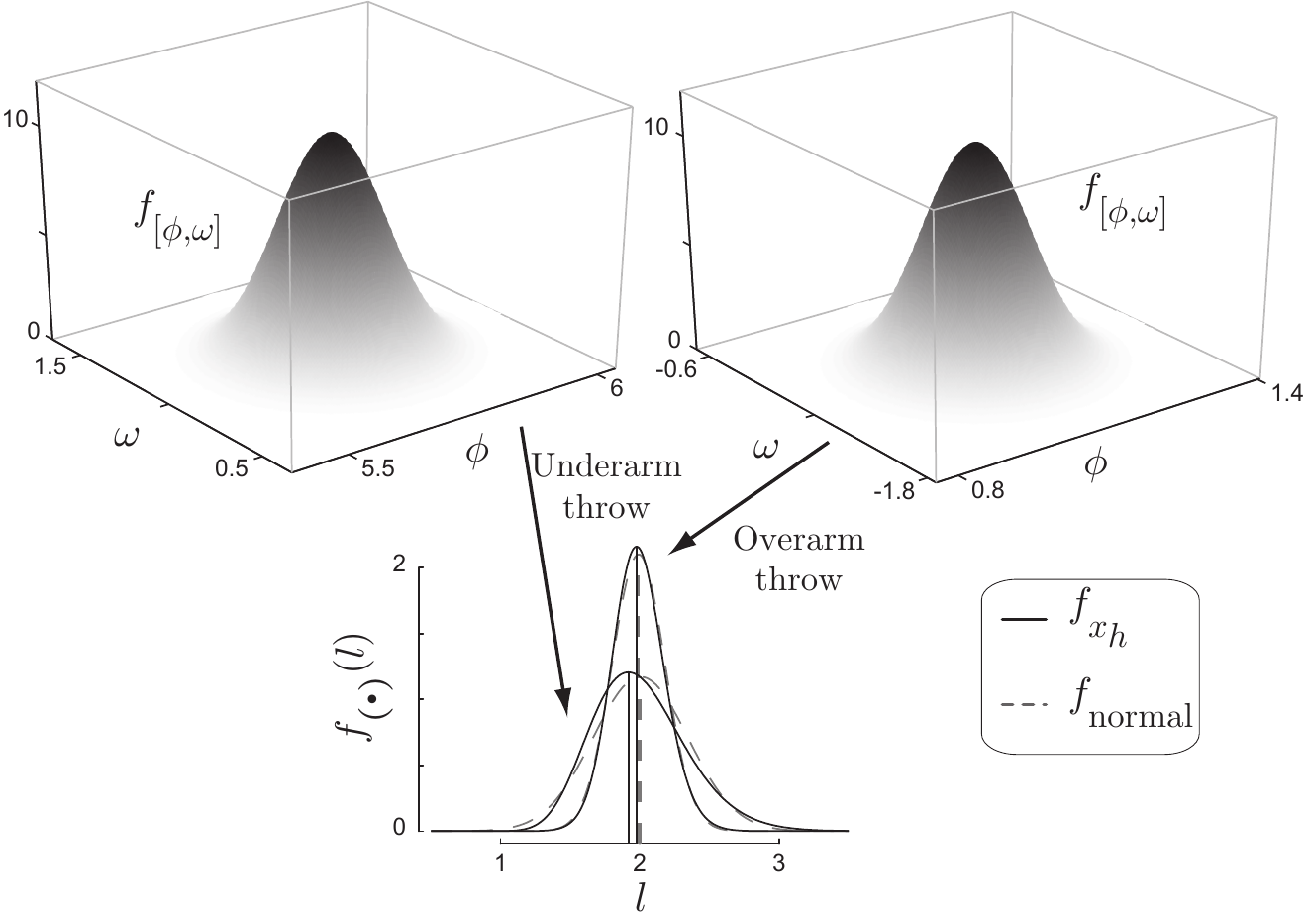}
	\caption{Propagating initial distributions with equal variance for hitting a target at $l=2$, $h=-1$. The underarm throw amplifies errors more than the overarm throw,  in agreement with the linear analysis. For 
	comparison, we superimpose a normal distribution (dashed curves) with mean and variance equal to the propagated distribution 
	(solid curves), together with the mode (solid vertical line) and mean (dashed vertical line). The overarm throw leads to a nearly 
	normal distribution where the projectile lands with mode (1.98) and mean (1.99) nearly equal to each other. The underarm throw in 
	contrast leads to a skewed distribution where the most likely outcome is an undershoot as seen from the mode (1.92) although the 
	mean (2.01) is nearly at the target ($l=2$). The input distributions are centred at ($\phi^\text{optim}_\text{over} = 1.082$, $
	\omega_0(\phi^\text{optim}_\text{over}) = -1.186$) for the overarm throw and at ($\phi^\text{optim}_\text{under} = 5.716$, $
	\omega_0(\phi^\text{optim}_\text{under}) = 1.008$) for the underarm throw, and we set standard deviations as $\sigma_\phi = 
	0.087$, $\sigma_\omega = 0.177$ for both throws. For a one metre long arm, these values expressed as mean $\pm$ standard 
	deviation are, overarm: $\phi=62\pm5^\circ, \abs{v}=3.72\pm0.56\, \text{m}/\text{s}\ (13.4\pm2\, \text{km}/\text{hr})$ and underarm: $\phi=-33\pm5^\circ, 
	\abs{v}=3.17\pm0.56\, \text{m}/\text{s}\ (11.4\pm2\, \text{km}/\text{hr})$.}
	\label{fig:planar_propagate_distributions_raster}
\end{figure}

Our numerical solutions show that the underarm throw (green solid curve in \figref{fig:planar_propagate_distributions_raster}) amplifies errors more 
than the overarm throw (red solid curve in \figref{fig:planar_propagate_distributions_raster}), as predicted by the linear amplification $\lambda$. Quantitatively as well, the ratio of overarm to underarm linear amplification is similar to the ratio of $\sigma$, the standard deviation of the output distribution $f_{x_{h}}$ (see table~\ref{tab:comparison}). However, the nonlinear calculation calculation shows that the output distribution for an underarm throw is skewed (\figref{fig:planar_propagate_distributions_raster}, table~\ref{tab:comparison}), something that cannot be found from a linear analysis. The implication of this skewed output is that the underarm throw could be perceived as a weaker throw because the mode of the distribution where the projectile is most likely to land, undershoots the target by 4\% in this example. In contrast, the overarm throw undershoots the target by only 1\%. We speculate that this undershoot, solely due to projectile dynamics, may partially underlie the common perception that an underarm throw is `weaker' than an overarm throw.

\begin{table}
\caption{Comparison of linear and nonlinear calculations for bin at $l=2$, $h=-1$.}
\label{tab:comparison}
\centering
\begin{tabular}{|c|c|c|c|c|c|}
\hline 
Linear & \multicolumn{5}{c|}{Nonlinear}\\
\hline 
\multirow{2}{*}{$\dfrac{\lambda_{\text{over}}}{\lambda_{\text{under}}}$} & \multirow{2}{*}{$\dfrac{\sigma_{\text{over}}}{\sigma_{\text{under}}}$} & \multicolumn{2}{c|}{Mean landing location} & \multicolumn{2}{c|}{Most likely landing location}\\
\cline{3-6}
 &  & Overarm & Underarm & Overarm & Underarm\\
\hline 
\multirow{2}{*}{0.69} & \multirow{2}{*}{0.56} & \multirow{2}{*}{1.99} & \multirow{2}{*}{2.01} & 1.98 & 1.92\\
&  &  &  & 1\% undershoot & 4\% undershoot\\
\hline
\end{tabular}
\longcaption{Linear: $\lambda(\phi^\text{optim}_\text{over})=1.35,\ \lambda(\phi^\text{optim}_\text{under})=1.96$; Nonlinear: $\sigma_\text{over} = 0.19,\ \sigma_\text{under} = 0.34$.}

\end{table}

\section{Shooting: Zero arm length}
\label{sec:ShootingZeroArmLength}
Our considerations so far use the length of the arm $R$ to set the length scale in the problem. We now look at the limit when this length scale becomes vanishingly small or for throwing to faraway targets where $L,H \gg R$. In this, the artillery limit, the problem becomes one of shooting a projectile at an angle $\theta$ with a linear velocity $V$.  Naturally therefore,  we only consider targets at the same height as the origin from where the projectile is launched.  The ratio of variability in $V$ to the variability in $\theta$, introduces a velocity scale $k = \delta v/\delta \theta$, where the quantities $\delta v$ and $\delta\theta$ are appropriate measures of variability. By 
expressing distances in units of $k^2/g$ and time in units of $k/g$, we obtain the scaled trajectory equations  for distance $x$, height $y$, as a function of time $t$, launch angle $\theta$ and launch velocity $v$,
\begin{subequations}\label{eqn:noarm_xyt}
\begin{align}
	x(t) &= t v\cos\theta \label{eqn:noarm_xt}\\
	y(t) &= t v \sin\theta - \frac{1}{2}t^2. \ \label{eqn:noarm_yt}
\end{align}
\end{subequations}
For exactly striking the target at a distance $l$ with the launch parameters $\theta, v$, the projectile lands at $x_0(\theta,v) = l = g L/k$. This gives the one-dimensional curve of launch parameters $(\theta_0,v_0(\theta_0))$:
\begin{equation}
	v_0(\theta) = \sqrt{\frac{l}{\sin 2\theta}}.
\label{eqn:noarm_v0}
\end{equation}
Performing a linearized error analysis in the neighbourhood of this curve  $(\theta_0,v_0(\theta_0))$ yields the Jacobian $J$ that maps small errors $\delta\theta$ and $\delta v$ to small errors $\delta x_0$:
\begin{subequations}\label{eqn:noarm_Jderivation}
\begin{align}
J(\theta_0) =	\begin{pmatrix}
								\pdiff{x_0}{\theta}& \pdiff{x_0}{v}
																\end{pmatrix}\bigg\vert_{\theta = \theta_0, v = 
																v_0(\theta_0)}
= \begin{pmatrix} 2v_0^2\cos 2\theta& 2v_0\sin 2\theta \end{pmatrix}
= 2 \begin{pmatrix} \frac{l\cos 2\theta}{\sin 2 \theta}& \sqrt{l\sin 2\theta} \end{pmatrix}. \label{eqn:noarm_J}
\end{align}
\end{subequations}
 The shooting angle most robust to small input noise corresponds to the $\theta$ where the only positive singular value is minimized.  Minimizing the square of the singular value 
\begin{equation}
	\zeta(\theta) = J^{\rm T}J= \frac{l^2 \cos^2 2\theta}{\sin^2 2\theta} + l\sin 2\theta \label{eqn:noarm_lambda}
\end{equation}
yields  
\begin{subequations}\label{eqn:noarm_solve_theta_star}
\begin{align}
	\frac{d\zeta}{d\theta}\bigg\vert_{\theta = \theta^\ast} = 0 \Rightarrow 4 l\cos2\theta^\ast - 8 l^2\frac{\cos 2\theta^\ast}{\sin^3 2\theta^\ast} &= 0\\
	\Rightarrow \cos 2\theta^\ast = 0,\ \text{or, for}\ 0<l<\frac{1}{2},\ \sin 2\theta^\ast &= \sqrt[3]{2 l} 
	\label{eqn:noarm_Dlambda_solutions}
\end{align}
\end{subequations}
We therefore find that for $l>1/2$, there is only one extremum at $\theta^\ast=\pi/4$, but three extrema co-exist for $0<l<1/2$ at $\theta^
\ast = \pi/4$, $\arcsin\sqrt[3]{2 l}/2$, and $\pi/2 - \arcsin\sqrt[3]{2 l}/2$. To determine if the solution \eqrefs{eqn:noarm_Dlambda_solutions} is a minimum or maximum, we evaluate  
\begin{subequations} \label{eqn:noarm_d2lambda}
\begin{align}
	\frac{d^2\zeta}{d\theta^2} &= -16 l \sin 2\theta + \frac{32 l^2 + 64 l^2 \cos^2 2\theta}{\sin^4 2\theta}\\
	\Rightarrow \frac{d^2\zeta}{d\theta^2}\bigg\vert_{\theta=\theta^\ast} &= \begin{cases}
	16 l (2 l - 1) &\text{when}\ \theta^\ast = \frac{\pi}{4},\ 0<l\\
	24(2 l)^\frac{2}{3}(1-(2 l)^\frac{2}{3}) &\text{when}\ \sin 2\theta^\ast = \sqrt[3]{2 l},\ 0<l<\frac{1}{2}								\end{cases} \label{eqn:noarm_DDlambda_solutions}
\end{align}
\end{subequations}
From the solutions in \eqrefs{eqn:noarm_DDlambda_solutions} we see that $\theta^\ast = \pi/4$ is a local minimum for $l>1/2$, a local 
maximum for $l<1/2$, and there is a pitchfork bifurcation at $l=1/2$. We also find that for $l<1/2$, when $\theta^\ast=\pi/4$ becomes a 
local maximum, there are two new branches that are a local minimum, with a cubic dependence on $l$ near $l=1/2$. 

The result that the 
optimal shooting angle ($\theta^\ast$) for targets with $l > 1/2$ is $\pi/4$ is well known, including cases where 
there is air drag \citep{Gablonsky2005}. However these past results however do not scale the equations like we do using the relative 
amount of noise in shooting velocity compared to shooting angle, and therefore do not identify the pitchfork bifurcation.  Thus $\theta = \pi/4$ becomes the worst possible choice for shooting when $l<1/2$, and there exist two 
symmetric branches of optimal shooting angles.

\begin{figure}
	\centering
		\includegraphics{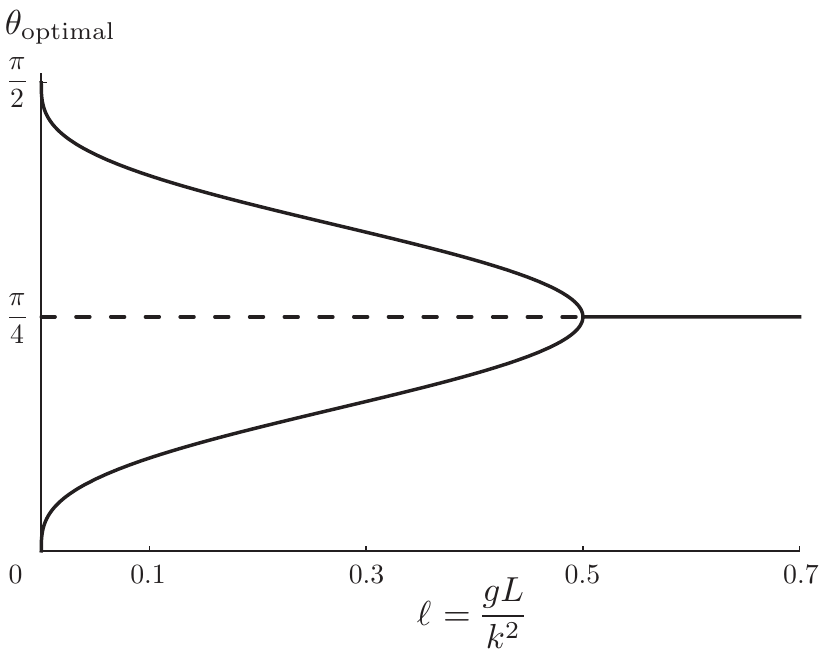}
	\caption{Bifurcation diagram for best shooting strategy based on a linearization in the neighborhood of all strategies that lead to an 
	accurate strike. Recall that $l = g L/k^2$ is the dimensionless distance to the target and $\theta_\text{optimal}$ is the angle to shoot 
	the target so that launching errors are amplified the least amount.}
	\label{fig:shooting_bifurcation}
\end{figure}

We attempt to provide physical intuition to this surprising result for the problem of shooting a drag-free projectile. In the limit of the relative 
noise in shooting velocity being much greater than that in the shooting angle, i.e.\ $k\gg 1 \Rightarrow l\approx 0$, the best  
 shooting angle corresponds to zero sensitivity to fluctuations in velocity, namely a high velocity shot  with $\theta =0$ or $\theta=\pi/2$. At the other extreme where $k\approx 0 \Rightarrow l\gg 1$, i.e.\  when the noise in shooting angle is much larger than 
in shooting velocity, $\pi/4$ is the optimal angle because that corresponds to a local maximum for projectile range, and hence has zero sensitivity to fluctuations in shooting angle. There is a bifurcation, and not a smooth continuous 
change between these two extreme cases because the problem is symmetric about $\pi/4$ and the optimal shooting angle depends 
continuously on $l$. The specific bifurcation point is determined by a trade off between the two sensitivities of the shooting range to 
shooting angle and velocity, respectively. For small perturbations $\epsilon$ near $\theta=\pi/4$, the sensitivity to shooting angle is $16 
l^2 \epsilon^2 + O(\epsilon^3)$ and the sensitivity to shooting velocity is $-8 l \epsilon^2 + 4 l + O(\epsilon^3)$. Therefore, in the 
neighbourhood of $\theta=\pi/4$ and for $l\le1/2$, the sensitivity to shooting velocity overpowers the sensitivity to shooting angle.

\section{Discussion}
\label{sec:Discussion}
Our calculations of optimal throwing strategies are based on a consideration of how infinitesimal noise in release parameters is propagated by the parabolic trajectory of a projectile. Past studies have found that greater error is permitted in throwing angle, than speed \citep{Gablonsky2005,Brancazio1981}. However, there is an inherent problem in how to compare two quantities that are dimensionally different. Past researchers either treat errors in release angle and velocity separately \citep{Dupuy2000}, or compare percentage errors \citep{Gablonsky2005,Brancazio1981}, quantities whose scaling is affected by the magnitude of throwing angle and velocity, how angle is measured and the relative weight of these measures.  By introducing a natural length scale, the arm length,  and making our formulation dimensionless, we can directly compare sensitivity to errors in angle and speed, thus avoiding multiobjective optimization that uses a heuristic weighting parameter. In our minimal model,  we ignored sensory feedback from the arm, covariance structure of the noise in input parameters,  dependence of noise on projectile mass, throwing speed, and so on. Nevertheless, we find a dependence of optimal strategy on target geometry, and a trade-off between speed and accuracy, which has until now been ascribed to signal dependent noise in muscles \citep{Fitts1954}. Our results agree with the prevalent observation in the literature that the slowest throw is most accurate \citep{Gablonsky2005,Brancazio1981,Tan1981,Okubo2006}. A minor modification that accounts for the distribution of angles associated with planning, allows our theory to make some testable predictions for the fraction of overarm throws used by humans in an experiment \citep{Venkadesana}.

We have also identified a non-trivial null-space of the mapping between errors in release parameters and errors in striking the target. This naturally suggests an additional control strategy to maximize accuracy of throwing: by tuning the covariance of input errors in angle and angular velocity to co-vary along the null-space of $J_\text{err}$ for small noise, or along the curve ($\phi_0,\omega_0(\phi_0)$) for large noise, one could mitigate output errors. In fact, there is some evidence that humans control the covariance of the projectile's launch angle and linear velocity in such a manner \citep{Dupuy2000,Kudo2000}. More generally, there is substantial evidence at the behavioural and neurophysiological levels that humans are adept at tuning covariance of noise across multiple joints / muscles \citep{Todorov2004,Valero-Cuevas2009}. However, because it is likely impossible to perfectly covary input errors, our calculations of maximal error amplification should predict strategies for accurate throwing even in the presence of structured noise. To see this, we note that geometrically, the effect of structured noise amounts to asking how an ellipse gets amplified by $J_\text{err}$. Even if the long axis of the ellipse is oriented parallel to the null-space of $J_\text{err}$, so long as it has any non-zero length along its minor axis, the non-zero singular value of $J_\text{err}$ specifies how input errors get amplified.

Finally an additional conclusion is that the most accurate throw is close to, but slightly faster than the slowest throw, an observation that has also been experimentally reported \citep{Freeston2007}. Moreover, we find that at higher speeds, there is a clear advantage for the overarm style. This is particularly salient in light of hunting and human evolution. Humans are significantly better (faster and more accurate) at overarm throwing than any other member of the animal kingdom \citep{Wood2007,Watson2001,Westergaard2000}. Given that hunting possibly played a role in the evolution of human form, could the physics of error propagation in throwing provide a partial explanation for the selective pressure for superior overarm throwing skills in humans ? 

\bibliographystyle{abbrvnat}

\begin{thebibliography}{29}
\providecommand{\natexlab}[1]{#1}
\providecommand{\url}[1]{\texttt{#1}}
\expandafter\ifx\csname urlstyle\endcsname\relax
  \providecommand{\doi}[1]{doi: #1}\else
  \providecommand{\doi}{doi: \begingroup \urlstyle{rm}\Url}\fi

\bibitem[Brancazio(1981)]{Brancazio1981}
P.~J. Brancazio.
\newblock {Physics of basketball}.
\newblock \emph{American Journal of Physics}, 49\penalty0 (4):\penalty0
  356--365, 1981.
\newblock \doi{10.1119/1.12511}.

\bibitem[Chowdhary and Challis(1999)]{Chowdhary1999}
A.~G. Chowdhary and J.~H. Challis.
\newblock {Timing accuracy in human throwing}.
\newblock \emph{J Theor Biol}, 201\penalty0 (4):\penalty0 219--229, 1999.

\bibitem[Cohen and Sternad(2009)]{Cohen2009}
R.~G. Cohen and D.~Sternad.
\newblock {Variability in motor learning: relocating, channeling and reducing
  noise.}
\newblock \emph{Exp Brain Res}, 193\penalty0 (1):\penalty0 69--83, Feb. 2009.
\newblock \doi{10.1007/s00221-008-1596-1}.

\bibitem[Dupuy et~al.(2000)Dupuy, Motte, and Ripoll]{Dupuy2000}
M.~A. Dupuy, D.~Motte, and H.~Ripoll.
\newblock {The regulation of release parameters in underarm precision
  throwing}.
\newblock \emph{J Sports Sci}, 18\penalty0 (6):\penalty0 375--382, 2000.

\bibitem[Etnyre(1998)]{Etnyre1998}
B.~R. Etnyre.
\newblock {Accuracy characteristics of throwing as a result of maximum force
  effort}.
\newblock \emph{Percept Mot Skills}, 86\penalty0 (3 Pt 2):\penalty0 1211--1217,
  1998.

\bibitem[Fitts(1954)]{Fitts1954}
P.~M. Fitts.
\newblock {The information capacity of the human motor system in controlling
  the amplitude of movement}.
\newblock \emph{J Exp Psychol}, 47\penalty0 (6):\penalty0 381--391, 1954.

\bibitem[Freeston et~al.(2007)Freeston, Ferdinands, and K]{Freeston2007}
J.~Freeston, R.~Ferdinands, and R.~K.
\newblock {Throwing velocity and accuracy in elite and sub-elite cricket
  players: A descriptive study}.
\newblock \emph{European Journal of Sport Science}, 7\penalty0 (4):\penalty0
  231--237, 2007.

\bibitem[Gablonsky and Lang(2005)]{Gablonsky2005}
J.~M. Gablonsky and A.~Lang.
\newblock {Modeling basketball free throws}.
\newblock \emph{SIAM Review}, 47\penalty0 (4):\penalty0 775--798, 2005.

\bibitem[Harris and Wolpert(1998)]{Harris1998}
C.~M. Harris and D.~M. Wolpert.
\newblock {Signal-dependent noise determines motor planning}.
\newblock \emph{Nature}, 394\penalty0 (6695):\penalty0 780--784, 1998.
\newblock \doi{10.1038/29528}.

\bibitem[Harris and Wolpert(2006)]{Harris2006}
C.~M. Harris and D.~M. Wolpert.
\newblock {The main sequence of saccades optimizes speed-accuracy trade-off}.
\newblock \emph{Biol Cybern}, 95\penalty0 (1):\penalty0 21--29, 2006.
\newblock \doi{10.1007/s00422-006-0064-x}.

\bibitem[Hopf(1934)]{Hopf1934}
E.~Hopf.
\newblock {On causality, statistics and probability}.
\newblock \emph{Journal of Mathematics and Physics}, 13:\penalty0 51--102,
  1934.

\bibitem[Hore et~al.(1996)Hore, Watts, and Tweed]{Hore1996}
J.~Hore, S.~Watts, and D.~Tweed.
\newblock {Errors in the control of joint rotations associated with
  inaccuracies in overarm throws}.
\newblock \emph{J Neurophysiol}, 75\penalty0 (3):\penalty0 1013--1025, 1996.

\bibitem[Kerr and Langolf(1977)]{Kerr1977}
B.~A. Kerr and G.~D. Langolf.
\newblock {Speed of aiming movements}.
\newblock \emph{The Quarterly Journal of Experimental Psychology}, 29\penalty0
  (3):\penalty0 475--481, 1977.

\bibitem[Kudo et~al.(2000)Kudo, Tsutsui, Ishikura, Ito, and Yamamoto]{Kudo2000}
K.~Kudo, S.~Tsutsui, T.~Ishikura, T.~Ito, and Y.~Yamamoto.
\newblock {Compensatory coordination of release parameters in a throwing task}.
\newblock \emph{J Mot Behav}, 32\penalty0 (4):\penalty0 337--345, 2000.

\bibitem[Mardia and Jupp(2000)]{Mardia2000}
K.~Mardia and P.~Jupp.
\newblock \emph{{Directional statistics}}.
\newblock Wiley New York, 2000.

\bibitem[Meyer et~al.(1990)Meyer, Smith, Kornblum, Abrams, and
  Wright]{Meyer1990}
D.~E. Meyer, J.~E.~K. Smith, S.~Kornblum, R.~A. Abrams, and C.~E. Wright.
\newblock {Speed-accuracy tradeoffs in aimed movements: Toward a theory of
  rapid voluntary action}.
\newblock \emph{Attention and Performance XIII}, pages 173--226, 1990.

\bibitem[Okubo and Hubbard(2006)]{Okubo2006}
H.~Okubo and M.~Hubbard.
\newblock {Dynamics of the basketball shot with application to the free throw}.
\newblock \emph{J Sports Sci}, 24\penalty0 (12):\penalty0 1303--1314, 2006.

\bibitem[Poincar\'{e}(1912)]{Poincare1912}
H.~Poincar\'{e}.
\newblock \emph{{Calcul des probabilit\'{e}s}}.
\newblock Gauthier-Villars, Paris, 1912.
\newblock URL \url{http://www.archive.org/details/calculdeprobabil00poinrich}.

\bibitem[Powls et~al.(1995)Powls, Botting, Cooke, and Marlow]{Powls1995}
A.~Powls, N.~Botting, R.~W. Cooke, and N.~Marlow.
\newblock {Motor impairment in children 12 to 13 years old with a birthweight
  of less than 1250 g}.
\newblock \emph{Arch Dis Child Fetal Neonatal Ed}, 73\penalty0 (2):\penalty0
  F62--6, 1995.

\bibitem[Smeets et~al.(2002)Smeets, Frens, and Brenner]{Smeets2002}
J.~B. Smeets, M.~A. Frens, and E.~Brenner.
\newblock {Throwing darts: timing is not the limiting factor}.
\newblock \emph{Exp Brain Res}, 144\penalty0 (2):\penalty0 268--274, 2002.

\bibitem[Tan and Miller(1981)]{Tan1981}
A.~Tan and G.~Miller.
\newblock {Kinematics of the free throw in basketball}.
\newblock \emph{American Journal of Physics}, 49\penalty0 (6):\penalty0
  542--544, 1981.

\bibitem[Timmann et~al.(1999)Timmann, Watts, and Hore]{Timmann1999}
D.~Timmann, S.~Watts, and J.~Hore.
\newblock {Failure of cerebellar patients to time finger opening precisely
  causes ball high-low inaccuracy in overarm throws}.
\newblock \emph{J Neurophysiol}, 82\penalty0 (1):\penalty0 103--114, 1999.

\bibitem[Todorov(2004)]{Todorov2004}
E.~Todorov.
\newblock {Optimality principles in sensorimotor control}.
\newblock \emph{Nat Neurosci}, 7\penalty0 (9):\penalty0 907--915, 2004.
\newblock \doi{10.1038/nn1309}.

\bibitem[Valero-Cuevas et~al.(2009)Valero-Cuevas, Venkadesan, and
  Todorov]{Valero-Cuevas2009}
F.~J. Valero-Cuevas, M.~Venkadesan, and E.~Todorov.
\newblock {Structured variability of muscle activations supports the minimal
  intervention principle of motor control}.
\newblock \emph{Journal of Neurophysiology}, 102\penalty0 (1):\penalty0 59--68,
  July 2009.
\newblock ISSN 0022-3077.
\newblock \doi{10.1152/jn.90324.2008}.

\bibitem[Venkadesan and Mahadevan(In Review)]{Venkadesana}
M.~Venkadesan and L.~Mahadevan.
\newblock {How to throw, accurately}.
\newblock \emph{Biology Letters}, In Review.

\bibitem[von Mises(1918)]{vonMises1918}
R.~von Mises.
\newblock {\"{U}ber die ``Ganzzahligkeit'' der Atomgewichte und verwandte
  Fragen}.
\newblock \emph{PhysikalischZe.}, 19:\penalty0 490--500, 1918.

\bibitem[Watson(2001)]{Watson2001}
N.~Watson.
\newblock {Sex differences in throwing: monkeys having a fling}.
\newblock \emph{Trends Cogn Sci}, 5\penalty0 (3):\penalty0 98--99, 2001.

\bibitem[Westergaard et~al.(2000)Westergaard, Liv, Haynie, and
  Suomi]{Westergaard2000}
G.~C. Westergaard, C.~Liv, M.~K. Haynie, and S.~J. Suomi.
\newblock {A comparative study of aimed throwing by monkeys and humans}.
\newblock \emph{Neuropsychologia}, 38\penalty0 (11):\penalty0 1511--1517, 2000.

\bibitem[Wood et~al.(2007)Wood, Glynn, and Hauser]{Wood2007}
J.~N. Wood, D.~D. Glynn, and M.~D. Hauser.
\newblock {The uniquely human capacity to throw evolved from a non-throwing
  primate: an evolutionary dissociation between action and perception}.
\newblock \emph{Biol Lett}, 3\penalty0 (4):\penalty0 360--364, 2007.

\end{thebibliography}

\appendix{Linearization for throwing with an arm}
\label{sec:LinearizationForThrowingWithAnArm}
When throwing with an arm (\secrange{sec:ModelOfTheThrowingTask}{sec:velocityAccuracyTradeOff}), the Jacobian of $x_h(\phi,
\omega)$ in the neighbourhood of the curve $(\phi,\omega_0(\phi))$, and its non-zero singular value is given by,
\begin{subequations}
\begin{align}
	J_\text{err}(\phi) &= \begin{pmatrix}
													\pdiff{x_h}{\phi}& \pdiff{x_h}{\omega}
												\end{pmatrix}\bigg\vert_{\phi,\omega_0(\phi)}\\
	\pdiff{x_h}{\phi}\bigg\vert_{\phi,\omega_0(\phi)} &= \frac{4 h l\cos\phi-4 (h+l\cot\phi)+\csc\phi
   \left(\left(2 l^2-1\right) \cos 2\phi+3\right)}{4 h\sin\phi
   +2 l \cos\phi+\cos 2\phi-3}\\
	\pdiff{x_h}{\omega}\bigg\vert_{\phi,\omega_0(\phi)} &= \frac{4\sqrt{2} \sin^2\phi
   (\csc\phi-l\cot\phi-h)^\frac{3}{2}}{4 h \sin (\phi )+2 l \cos (\phi )+\cos (2 \phi
   )-3}
\end{align}
\end{subequations}
The non-zero singular value $\lambda(\phi)$ of $J_\text{err}(\phi)$ is given by,
\begin{align}
	\lambda^2(\phi) &= \left(\pdiff{x_h}{\phi}\bigg\vert_{\phi,\omega_0(\phi)}\right)^2 + \left(\pdiff{x_h}{\omega}\bigg\vert_{\phi,
	\omega_0(\phi)}\right)^2 \label{eqn:singular value formula}
\end{align}

\appendix{Feasible arm angles}
\label{sec:FeasibleArmAngles}

\begin{figure}[!h]
	\centering
		\includegraphics{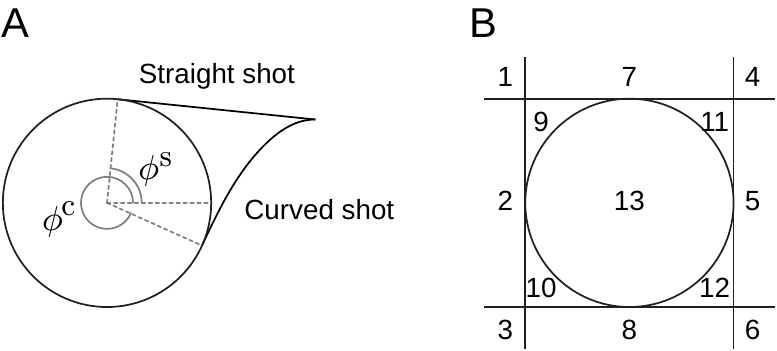}
	\caption{The individual cases for identifying the feasible arm angles in order to be able to strike a planar target. \textbf{a.} Two 
	qualitatively different shallow shots to hit the target depending on whether the target is above or below the point of release. 
	\textbf{b.} Enumeration of different regions of target location where domains of $\phi$ are calculated for $x_h(\phi,\omega_0(\phi))$ 
	to be a real number.}
	\label{fig:feasible_arm_angles}
\end{figure}

To calculate the arm angles where it is possible to hit the target, we first observe that there are two qualitatively different shallow shots 
depending on the height of the target relative to the release point (\figref{fig:feasible_arm_angles}a). We refer to the straight-line shot with 
infinite velocity by $\phi^\text{s}_{(\centerdot)}$ and the curved shallow shot which is a parabola passing through the target with its apex 
at the target, and tangential to the unit circle by $\phi^\text{c}_{(\centerdot)}$. The subscript will denote the throwing style (`u' or `o'). 
These angles are give by,
\begin{subequations}\label{eqn:shallow and curved shot limits}
\begin{align}
	c(\alpha) &= (1+l)\tan^4\frac{\alpha}{2} - 4 h\tan^3\frac{\alpha}{2} + 6\tan^2\frac{\alpha}{2} - 4 h\tan\frac{\alpha}{2} + (1-l) 
	\label{eqn:curved shot quartic equation}\\
	\phi^\text{c}_\text{o} &= \begin{cases}
											\max\alpha &\text{where}\ \alpha\ \text{is a real root of}\ c(\alpha) \text{if}\ l<1\\
											\min\alpha &\text{if}\ l>1
										\end{cases}\label{eqn:curved shot overarm}\\
	\phi^\text{c}_\text{u} &= \begin{cases}
											\min\alpha &\text{if}\ l<1\\
											\max\alpha &\text{if}\ l>1
										\end{cases}\label{eqn:curved shot underarm}\\
	\phi^\text{s}_\text{o} &= \arctan2(h,l) + \arccos\frac{1}{d}\label{eqn:straight shot overarm}\\
	\phi^\text{s}_\text{u} &= \arctan2(h,l) -	\arccos\frac{1}{d}\label{eqn:straight shot underarm}\\
	\arctan2(h,l) &= \arg(l+i h)\label{eqn:arctan2}\\
	d &= \sqrt{l^2+h^2}\label{eqn:formula for distance d}
\end{align}
\end{subequations}
\eqref{eqn:curved shot quartic equation} has only two real roots in the interval $[0,2\pi]$. The domains of feasible $\phi$ so that $x_h(\phi,
\omega_0(\phi))$ is a real number are given by,
\begin{align*}
	1:\ &l<-1, h\ge1 &\Phi &= (0, \phi_\text{u}^\text{c}] \cup (\pi, \phi_\text{o}^\text{c}]\\
	2:\ &l<-1, \abs{h}<1 &\Phi &= (0, \phi_\text{u}^\text{s}) \cup (\pi, \phi_\text{o}^\text{c}]\\
	3:\ &l<-1, h\le-1 &\Phi &= (0, \phi_\text{u}^\text{s}) \cup (\pi, \phi_\text{o}^\text{s})\\
	4:\ &l>1, h\ge1 &\Phi &= [\phi_\text{u}^\text{c}, 2\pi) \cup [\phi_\text{o}^\text{c}, \pi)\\
	5:\ &l>1, \abs{h}<1 &\Phi &= [\phi_\text{u}^\text{c}, 2\pi) \cup (\phi_\text{o}^\text{s}, \pi)\\
	6:\ &l>1, h\le-1 &\Phi &= (\phi_\text{u}^\text{s}, 2\pi) \cup (\phi_\text{o}^\text{s}, \pi)\\
	7:\ &\abs{l}<1, h>1 &\Phi &= (0, \phi_\text{u}^\text{c}] \cup [\phi_\text{o}^\text{c}, \pi)\\
	8:\ &\abs{l}<1, h<-1 &\Phi &= \left((0, \psi] \cup (\phi_\text{u}^\text{s},2\pi-\psi]\right) \cup \left([\psi,\pi) \cup [2\pi-\psi, \phi_\text{o}^
	\text{s})\right)\\
	9:\ &-1<l<0, 0<h\le1 &\Phi &= (0, \phi_\text{u}^\text{s}) \cup [\phi_\text{o}^\text{c}, \pi)\\
	10:\ &-1<l<0, -1\le h<0 &\Phi &= \left((0, \psi] \cup (\phi_\text{u}^\text{s},2\pi-\psi]\right) \cup \left([\psi,\pi) \cup [2\pi-\psi, \phi_\text{o}^
	\text{c}]\right)\\
	11:\ &0<l<1, 0<h\le1 &\Phi &= (0, \phi_\text{u}^\text{c}] \cup (\phi_\text{o}^\text{s}, \pi)\\
	12:\ &0<l<1, -1\le h<0 &\Phi &= \left((0, \psi] \cup [\phi_\text{u}^\text{c},2\pi-\psi]\right) \cup \left([\psi,\pi) \cup [2\pi-\psi, \phi_\text{o}^
	\text{s})\right)\\
	13:\ &l^2+h^2\le1 &\Phi &= (0, \psi] \cup [\psi, \pi)
\end{align*}
where $\psi = \arccos(l)$, and the domains are listed in the format (underarm) $\cup$ (overarm).

\end{document}